\documentclass[structabstract]{aa}
\usepackage{graphicx}
\usepackage{txfonts}
\begin{document}
   \title{ Notes on disentangling of spectra}

   \subtitle{II. Intrinsic line-profile variability
     due to Cepheid pulsations\thanks{This study uses
           the spectra from the Ond\v{r}ejov 2-m telescope.}
            }
   \author{P. Hadrava\inst{1}
           \and
           M. \v{S}lechta\inst{2}
           \and
           P. \v{S}koda\inst{2}
          }

   \institute{Astronomical~Institute, Academy~of~Sciences,
              Bo\v{c}n\'{\i}~II~1401, CZ~-~141~31~Praha~4,
              Czech~Republic\\
              \email{had@sunstel.asu.cas.cz}
              \and
              Astronomical~Institute, Academy~of~Sciences,
              Fri\v{c}ova 298, CZ~-~251~65~Ond\v{r}ejov,
              Czech~Republic\\
              \email{slechta@sunstel.asu.cas.cz, skoda@sunstel.asu.cas.cz}
             }

   \date{Received May 22, 2009; accepted August 2, 2009}


  \abstract
  {The determination of pulsation velocities from observed spectra of
  Cepheids is needed for the Baade-Wesselink calibration of these
  primary distance markers.}
  {The applicability of the Fourier-disentangling technique
  for the determination of pulsation velocities of Cepheids and other
  pulsating stars is studied.}
  {The KOREL-code was modified to enable fitting of free parameters
  of a prescribed line-profile broadening function corresponding to
  the radial pulsations of the stellar atmosphere. It was applied to
  spectra of $\delta$~Cep in the H-alpha region observed with the Ond\v{r}ejov
  2-m telescope.}
  {The telluric lines were removed using template-constrained
  disentangling, phase-locked variations of line-strengths were measured
  and the curves of pulsational velocities obtained for several
  spectral lines. It is shown that the amplitude and phase of the
  velocities and line-strength variations depend on the depth of line
  formation and the excitation potential.}
   {The disentangling of pulsations in the Cepheid spectra may be
   used for distance determination.}


   \maketitle
%

\section{Introduction}

The method of Fourier disentangling of spectra was developed by
Hadrava (1995) from the method of cross-correlation (cf., e.g., Hill
1993) to decompose, from a series of observed spectra of multiple
stars, contributions of the individual components and to
simultaneously
find orbital parameters, or, more generally, to fit free parameters
of the physics governing the Doppler shifts and line-profile variations.
One of the advantages of disentangling compared to cross-correlation
is that it does not require a template spectrum of a star with
similar spectral type or a model atmosphere.

 Zucker and Mazeh (2006) introduced their method of Template
Independent RAdial-VELocity measurement (TIRAVEL) for single-lined
spectroscopic binaries (SB1). Their method uses each exposure
from a series of observations as a template for cross-correlation with
all other exposures. Zucker and Mazeh mentioned that the KOREL-code
(cf., e.g., Hadrava 2004b) for Fourier disentangling is also
template-independent. Nevertheless, they advocated use
of their TIRAVEL based on the alleged advantage that it
assumes the individual radial velocities to be free variables
not bounded by an orbital motion. However, 
KOREL 
enables the convergence of individual radial velocities
of any component also. Moreover, KOREL can do this not only for SB1,
which is an extremely simple case, but also for
two or more components. The option of free velocities is rarely used,
because in practice it is more advantageous to take the orbital motion
of multiple stars into account and to solve directly for the orbital
parameters. Zucker and Mazeh mentioned the case of a third component
as an example of when free radial velocities are needed. However, just in this
case the solving for orbital parameters of both the close and the wide orbit
(taking into account also the light-time effect)
simultaneously with the determination of radial velocities is advantageous,
because it checks the consistency of the possibly small perturbation with
the source spectra better than a two-step procedure of determination
and subsequent solution of the radial-velocity curve.

 One reason to solve in some cases for individual radial velocities 
independent of
orbital motion is technical, e.g. in the case of an unreliable wavelength
scale of the observed spectra, as it has been done in the study by Yan et al.
(2008). However, a more important reason is seen
in cases when the observed wavelength shifts of spectral lines are not
due to overall orbital motion of the star but due to some other effects.
One such case is the Doppler shift caused by pulsations of the stars.
The use of Fourier disentangling in the spectroscopic studies of
pulsating stars has been outlined by Hadrava (2004a,b). Here we shall
demonstrate this method in practice in the case of the star $\delta$~Cep.

 The study of either radially or non-radially pulsating stars is
important as a clue to probe the inner structure of stars. In addition,
the period-luminosity (PL) relation of Cepheids and some other radial
pulsators is used as one of the few primary methods for distance
determination. One needs to calibrate this PL-relation, and that
can be done by the Baade-Wesselink method (Baade 1926, Wesselink 1946).
In this method, the spectroscopically measured pulsation velocity is
integrated over the course of the period to yield the changes of stellar
radius, which, combined with photometric or interferometric variations, may
reveal the distance to the star. However, the spectroscopic measurement
is complicated by the line-profile variations caused not only by the radial
motion of the stellar atmosphere (which must be integrated over the visible
part of the stellar disc projected locally to the line of sight) but also due
to changes of other physical conditions (e.g. the temperature and density)
in line-forming regions of different lines. Consequently, the radial
velocities $v_{r}$ measured by different methods reflect only indirectly the
instantaneous pulsation velocity $v_{p}$ of the stellar surface. It is thus
used to introduce the so called projection factor $p\equiv v_{p}/v_{r}$,
which can be estimated either theoretically (e.g. Nardetto et al. 2004)
or observationally (Nardetto et al. 2008 and citations therein). An
alternative method directly matching the Doppler shifts and asymmetries
of spectral lines with a proper model of line-profiles has been introduced
by Gray and Stevenson (2007). Their approach is, in principle, equivalent
to the above mentioned disentangling of pulsations (Hadrava 2004a,b).

 We summarize the method of pulsation disentangling and its
use in Sect.~\ref{Theor} and we test it on our observations of
the $\delta$~Cep in Sect.~\ref{Disent}. In Sect.~\ref{Concl}
we discuss possibilities of further development of the method.


\section{Disentangling of radial pulsations}\label{Theor}

The line-profile variations (LPVs) caused by motion of the stellar
atmosphere are often modelled by integrating Doppler shifted spectra
of unperturbed (mostly plane-parallel) model atmospheres over the stellar
surface. This approximation is used for study of the rotational broadening
as well as LPVs due to radial or non-radial pulsations even though it is
obvious that the motion may change the structure of the atmosphere and hence
also the radiative transfer and the line formation in them. In the case of
Cepheids, it is observed that different lines formed in different layers of
the atmosphere have slightly different phase-dependence of LPVs
(cf. Breitfellner and Gillet 1993, Butler 1993 etc.), which reflects
(besides other effects) the changes of velocity gradients within the
atmosphere. The basic idea of the Baade-Wesselink method also assumes
that the spectroscopic and photometric or interferometric variations are
caused by pulsations of the same stellar surface. However, the dilution
of the atmosphere and its temperature variations and the presence
of stellar winds in these stars implies that the motion of layers with
given optical depths may differ from the local velocity of the gas, which
then influences the profiles and shifts of the spectral lines.

 A safe way (outlined and followed, e.g., by A. Fokin, 1991, 2003 etc.)
to treat all these effects properly would be to construct physically
self-consistent models of atmospheres of pulsating stars, to calculate
the observable quantities (spectra, light-curves, visibility functions)
and to match the values of free parameters (including the distance) to
the real data. However, the modelling itself is still computationally
very demanding, not to speak of the inverse problem of solving for
the free parameters. It is thus worth simultaneously following
an alternative way of fitting the observations by simplified models,
which take into account the most important effects and could be
modified if a discrepancy with respect to self-consistent models or with
respect to observations were found.

 Let us assume now that the observed spectrum $I(x,t)$ (as a function
of the logarithmic wavelength $x$ and time $t$) is given by the integral
\begin{equation}\label{comp2}
 I(x,t)= \int_{s}\mu I(x,s,\mu,t)\ast \delta(x-v(s,t)) d^{2}s
\end{equation}
over the visible part (i.e. where the directional cosine $\mu>0$) of
the stellar surface $s$. 
If the specific intensity (in the rest frame of the moving atmosphere) is
independent of $s$ and we expand it into a power series of $\mu$
\begin{equation}\label{comp1}
  I(x,s,\mu,t)=\sum_{k}I^{k}(x,t)\mu^{k}\; ,
\end{equation}
the spectrum is given by
\begin{equation}\label{comp}
 I(x,t)= \sum_{k}I^{k}(x,t)\ast\Delta^{k}(x,t)\; ,
\end{equation}
where
\begin{equation}\label{broad}
 \Delta^{k}(x,t)= \int_{s}\mu^{k+1}\delta(x-v(s,t)) d^{2}s\; .
\end{equation}
For purely radial pulsations synchronous on the whole stellar surface,
the local radial velocity $v(s,t)=\mu v_{p}(t)$ is the projection of
the instantaneous speed $v_{p}$ of the pulsation. The broadening
functions for individual modes of the limb darkening then read
\begin{equation}\label{broadk}
 \Delta^{k}(x,t)=\frac{2\pi R^{2}}{v_{p}^{k+2}}\left[x^{k+1}\right]_{0}^{v_p}\; ,
\end{equation}
where $R=R(t)$ is the instantaneous radius of the star. The Doppler shift
which can be measured by the first moment is given by the mean value of $x$
of the broadening and it reads
\begin{equation}\label{vr}
 v_{r}\equiv \frac{\int x \Delta^{k}(x,t)dx}{\int \Delta^{k}dx}
  = \frac{k+2}{k+3}v_{p}\; .
\end{equation}

 In agreement with Getting (1934), the projection factor $p$ is thus
$\frac{3}{2}$ for lines without limb darkening and $\frac{4}{3}$ with
linear limb darkening equal to 1. For higher order terms of the series
given by Eq.~(\ref{comp1}), the projection factor decreases toward
the limiting value 1, which corresponds to the broadening function
\begin{equation}\label{broadinf}
 \Delta^{\infty}(x,t)\sim\delta(x-v_{p}(t))\; ,
\end{equation}
i.e. to an absorption occurring in the centre of the stellar disc only.
This calculation is valid exactly for radial-velocity measurement
by the moment method only. For other methods, like the deepest point
of the profile, bisectors, the fit by two semi-Gaussian curves
or the cross-correlation method,
the result of radial-velocity measurement of the asymmetric lines is
less certain. It also depends on the instrumental broadening. Because
the standard Fourier disentangling (which assumes the broadening function
given by Eq.~(\ref{broadinf}) only) is a handy method of radial-velocity
measurement, it is worth investigating its properties (and its
$p$-factor in particular) when applied to Cepheids, or modifying
it by taking into account the proper broadening function so that it
will avoid the $p$-factor completely and will directly provide
the pulsational velocities.

 It is generally understood that limb darkening is crucial for
the asymmetry of the lines and thus also for setting the projection
factor.
However, determination of its proper value is still a problem (cf.
Monta$\tilde{\rm n}$\'{e}s Rodriguez and Jeffery 2001, Marengo et al.
2002). The limb-darkening corresponding to the radiation in continuum is
sometimes accepted also in the lines, arguing that it ``varies slowly
with the wavelength and can be taken to be constant over the span of
a spectral line" (Gray 2005, p. 436). However, the spectral flux in
continuum also varies slowly with the wavelength, but it cannot be
taken to be constant over any line because it is just its fast
variation across the line-profile that is seen as the spectral
lines. A simple
model of radiative transfer in stellar atmospheres reveals that weak
lines are usually dominated by the central part of the visible stellar
disc and hence the differences between the radiation in the lines and
in continuum has the limb darkening close to 1.0 (cf. Hadrava 1997).
This is confirmed by the observed behaviour of the line strengths
during eclipses and it is also consistent with the values $p\simeq 1.3$
of the projection factor in Cepheids. However, detailed non-LTE models
of stellar atmospheres show that the limb darkening within strong lines
(e.g. Balmer lines) is more complex (cf. Hadrava and Kub\'{a}t 2003).
It can be represented as a superposition of a part without limb darkening,
a part with linear darkening, as well as of higher order terms,
which also may include an emission (especially at the outer edges of
the stellar disc if the spherical instead of the plane-parallel symmetry
is taken into account). We shall thus investigate separately the
cases of simulated pulsationally broadened lines without any limb
darkening ($k=0$) and with linearly proportional broadening ($k=1$).
We deconvolve these profiles (and in the next Section also real
observed spectra) by disentangling with broadening functions for the
same cases ($k=0, 1$) and by the standard disentangling (i.e. $k=\infty$).

 The generalization of disentangling for line-profile variability has
been described by Hadrava (1997 and 1998 for the line-strength variability and
the changes in shape of line-profile, respectively; unfortunately, the formulae
were incorrectly processed in the print of the later paper). The explicit form
of the pulsational broadening and its Fourier transform was given for
the case of limb-darkening equal to 1 by Hadrava (2004a,b). The Fourier
transform (from the space of functions of the variable $x$ to functions
of $y$) of this profile (Eq.~(\ref{broadk}) for $k=1$) has the form
\begin{equation}\label{fbroadp1}
 \tilde{\Delta}^{1}(y,t)=\frac{2\pi i R^{2}}{y^{3}v_{p}^{3}(t)}
  \left[\exp(iyv_{p})(2-2iyv_{p}-y^{2}v_{p}^{2})-2\right]\; .
\end{equation}
Here we need also the case with zero limb-darkening ($k=0$), for
which the Fourier transform reads
\begin{equation}\label{fbroadp0}
 \tilde{\Delta}^{0}(y,t)=\frac{2\pi R^{2}}{y^{2}v_{p}^{2}(t)}
  \left[\exp(iyv_{p})(1-iyv_{p})-1\right]\; ,
\end{equation}
and also the standard disentangling, for which $\tilde{\Delta}^{\infty}(y,t)
\simeq\exp(iyv_{p})$. The changes of $R(t)$ in these formulae influence
the flux in the lines in absolute units proportionally also to the
variations of the flux in continuum. Hence, they should not affect the
line-strengths in the rectified spectrum (i.e. normalized to the continuum),
unless there is also light from a binary companion
or a background star present in the spectrum. 
However, because the changes of the effective
temperature and other parameters of the stellar atmosphere in the course
of the pulsational cycle generally affect the line-strengths, we substitute
the term $2\pi R^2\equiv(k+2)s(t)$, where $s(t)$ is a multiplicative
line-strength factor given by the integral of Eq.~(\ref{broadk}) over $x$
(which scales the broadening normalized to a unit integral).

\begin{figure}
 \centering
 \includegraphics[width=8.6cm]{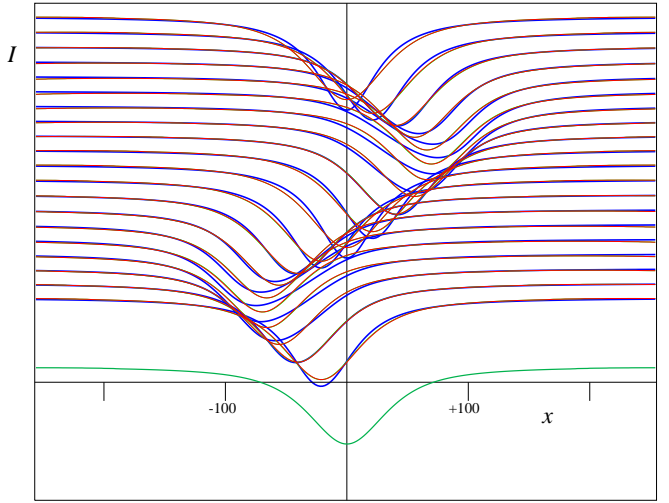}
   \setlength{\unitlength}{1mm}
   \put(-17,12){$x$}
   \put(-88,60){$I$}
 \caption{ Standard disentangling (with $\Delta^{\infty}_{dis}$) of
 simulated Lorentzian profiles broadened by radial pulsations (with
 $\Delta^{0}_{sim}$, $q=0.3$). The simulated profiles are drawn by
 thick lines (blue in the electronic version), their fits
 using disentangling by thin (red) lines and the mean disentangled
 profile by the bottom (green) line.}
 \label{obr1}
\end{figure}

 We have calculated several sets of simulated data by direct integration
in the $x$-space to always convolve one fixed Lorentzian line-profile
($\phi(x)\sim((x-x_0)^2+\gamma^2)^{-1}$) with
broadening functions either $\Delta^{0}$ or $\Delta^{1}$ for instantaneous
values of velocity $v_{p}(t)$ of a harmonic (i.e. sinusoidal) pulsation
at 20 values of $t$ uniformly covering the period of the pulsation.
The Lorentzian profile was chosen because it contains both a narrow
core and wide wings. An example of such Lorentzian profiles with
the intrinsic semi-halfwidth $\gamma$ corresponding to 30km/s broadened
by $\Delta^{0}$ with a semi-amplitude of the pulsational velocity
$K_{sim}$ equal to 100km/s is presented in Fig.~\ref{obr1}. (The
step of 100km/s is marked by the ticks on the $x$-axis.)
It can be seen that the profiles (cf. the thick lines with offsets
proportional to the phase) have the highest asymmetry at extremes
of the pulsational velocity, the high-velocity wing always being
steeper than the wider low-velocity wings reaching the rest wavelength
of the line. Qualitatively the same feature also results if
the pulsational broadening is applied to non-Lorentzian line
profiles and is in agreement with the line-profile variations
observed in Cepheids (cf., e.g., Breitfellner and Gillet 1993, Nardetto
et al. 2006, Gray and Stevenson 2007). It is used to characterize such
profiles quantitatively by bi-Gaussian fits (cf. Nardetto et al. 2006),
which, however, are not physically substantiated.

 Figure~\ref{obr1} also shows the disentangling of these simulated
profiles by the standard KOREL disentangling (i.e. $\Delta^{\infty}$).
For spectra containing lines of two or more component stars,
the disentangling simultaneously decomposes them and fits
the orbital parameters, while in the present case it leads
to a simpler problem of fitting all the input spectra by a mean profile
(which is drawn by the lowermost line) scaled in the line-strength and
shifted in $x$. The best fits to the individual input profiles are overplotted
in Fig.~\ref{obr1} by the thin lines. The disentangled mean profile is
symmetric here (which need not be exactly the case if the pulsational
velocity includes some higher overtones in addition to the basic
sinusoidal mode, or if the spectra do not cover the period uniformly).
Yet, it reproduces the input profiles relatively well, so that for real
data the difference between the true profiles and the fit by a simplified
model may be hidden in the noise. The coincidence is even better for
input profiles broadened by $\Delta^{1}$ and disentangled by
$\Delta^{\infty}$. In both cases the coincidence is worse if the amplitude
of the pulsational velocity exceeds more the intrinsic width of the line.
The disentangled profile is wider than the intrinsic profile
(this is well seen e.g. in the uppermost input line, for which
$v_{p}=0$), whenever the disentangling is performed using a smaller
broadening (i.e. $\Delta^{k}$ with higher $k$) than the broadening
used for the simulation of the data. If the input profiles are
disentangled with the same broadening for which they were
simulated, their fit as well as the reconstruction of the
intrinsic profile is perfect within the numerical precision
of the simulation.
The disentangled line-strength factors are higher (up to nearly
+0.08) for phases with low pulsational velocities and smaller
(nearly $-0.08$) at extremes of the velocity.

 The disentangling of this and other simulated datasets is
performed here with velocities bound to a sinusoidal pulsation,
which was also assumed in the creation of the data. This is done
by formally assuming that the velocity obeys a circular motion.
The ratio of the true amplitude $K_{sim}$ of the velocity chosen for
the simulation and its disentangled value $K_{dis}$ gives a mean value
of the projection factor $p$ for a particular combination of
the broadening functions. The values of pulsational velocities
which give the best fit of the mean disentangled profile with
the individual input profiles are also calculated and they
reveal that the projection factor is slightly phase-dependent.
However, the differences are very small -- in the particular
case shown in Fig.~\ref{obr1} the free radial velocity is about
0.62km/s (i.e. nearly 1\%) higher than the harmonic one at their
extremes and about 0.47km/s lower at medium (non-zero) values
of velocity. The profiles reconstructed with both velocities
are drawn in Fig.~\ref{obr1} by the thin lines, but they are not
distinguishable within the precision of the graphics. We thus neglect
these differences which are of the order of the precision of the
computation and we give the mean values of projection factors
$p=K_{sim}/K_{dis}$ only for several combinations of broadening
functions used for the simulation ($\Delta^{k}_{sim}$, $k=0,1$)
and the disentangling ($\Delta^{k}_{dis}$, $k=0,1,\infty$) and
the velocity-amplitude to line-width ratios $q=\gamma/K_{sim}$ in
Table~\ref{tab1}.


\begin{table}
 \caption[]{Projection factors for simulated line-profiles}
  \label{tab1}
 $$
   \begin{array}{rrrrrrrrr}
     \hline
     \noalign{\smallskip}
 && &\Delta^{0}_{sim} &  && &\Delta^{1}_{sim} &  \\
     \noalign{\smallskip}
 q && \Delta^{0}_{dis} & \Delta^{1}_{dis} & \Delta^{\infty}_{dis}
   && \Delta^{0}_{dis} & \Delta^{1}_{dis} & \Delta^{\infty}_{dis} \\
      \noalign{\smallskip}
      \hline
      \noalign{\smallskip}
 0.1 && 0.9999 & 1.0718 & 1.3942 && 0.9426 & 0.9999 & 1.2556 \\
 0.3 && 0.9999 & 1.1074 & 1.4624 && 0.9100 & 0.9999 & 1.3064 \\
 1.0 && 1.0001 & 1.1229 & 1.4945 && 0.8918 & 1.0001 & 1.3297 \\
 3.0 && 1.0013 & 1.1262 & 1.5012 && 0.8905 & 1.0015 & 1.3348 \\
 10. && 1.0006 & 1.1253 & 1.5007 && 0.8895 & 1.0003 & 1.3341 \\
\noalign{\smallskip}
      \hline
   \end{array}
     $$
\end{table}

 It can be seen from these results that for lines broader than their
Doppler shift ($q=\gamma/K_{sim}>1$) the standard disentangling (with
$\Delta^{\infty}_{dis}$) yields $p$-factors agreeing within the
numerical errors (which are of the order $10^{-3}$ here) with the
moment method of the radial-velocity measurement, i.e. $\frac{4}{3}$
or $\frac{3}{2}$ for lines with unit ($\Delta^{1}_{sim}$) or zero
($\Delta^{0}_{sim}$) limb darkening. For lines with intrinsic widths
$\gamma$ smaller than the pulsationally induced shifts and asymmetries,
the standard disentangling is more sensitive to the position of the
deeper parts of the profile and the $p$-factor decreases slightly closer
to the value 1. The disentangling with the proper limb-darkening in
the line (i.e. $\Delta^{1}_{dis}$ for $\Delta^{1}_{sim}$ and
$\Delta^{0}_{dis}$ for $\Delta^{0}_{sim}$) has $p=1$, it means that
it directly provides the pulsational velocities and it is desirable
to use it for the Baade-Wesselink calibration. The use of improper
broadening functions (i.e. $\Delta^{1}_{dis}$ for $\Delta^{0}_{sim}$
or vice versa), however, results in an error of about 10\% in
radial velocity. We thus need either to find
the proper limb darkening across the line-profiles from detailed
model atmospheres, or to distinguish which model fits the observed
line-profiles better.


\section{Disentangling of the observed spectra}\label{Disent}

To test the disentangling of pulsations on real data we started
spectroscopic observations of $\delta$~Cep using the 700-mm camera
of the spectrograph in the Coud\'{e} focus of the Ond\v{r}ejov 2-m
telescope equipped with LN2-cooled SITe CCD detector ST-005A
(2030 $\times$ 800 15-$\mu$m pixels).
Sixty nine medium-resolution
spectra ($R\sim$ 13 000) with a linear dispersion of 17 \AA/mm (0.25\AA/pix)
in H$\alpha$ region (6250--6770 \AA) obtained between August 19, 2008 and
April 16, 2009 (mostly by M. \v{S}lechta) are used in this study.
See on-line Table~\ref{tab3} or \ref{tab4} for the journal
of observations.

The spectra were reduced in IRAF\footnote{IRAF is distributed by
 the National Optical Astronomy Observatory,
 which is operated by the Association of Universities for Research in
 Astronomy (AURA), Inc., under cooperative agreement with the National
 Science Foundation}
using the standard packages {\tt ccdproc}, {\tt doslit} and {\tt rv}
(for more details of the processing see \v{S}koda and \v{S}lechta, 2002).

The spectral region around the H$\alpha$-line contains many atmospheric
water-vapour lines which complicate the measurement of stellar spectra
by standard methods (cf., e.g., Kiss and Vink\'{o} 2000). However, 
Fourier disentangling with variable line strengths is not only suitable
to remove the telluric lines (cf. Hadrava 1997, 2004a,b, 2006a), but at
the same time it enables one to use them for an additional check or
correction of the wavelength scale similarly to their use in 
classical methods (cf., e.g., Butler and Bell 1997). As the first step
of the disentangling we chose the spectral region 6511--6521~{\AA}
sampled in 1024 bins (i.e. with a step in radial velocity of 0.45km/s per bin)
to find the line-strength coefficients for the telluric lines. The Doppler
shifts of the telluric lines with respect to the heliocentric wavelength-scale
were calculated in the Keplerian approximation of the annual motion which
is provided by the PREKOR-code from the coordinates of the target star (cf.
Hadrava 2004b) and the telluric lines were disentangled by the standard
disentangling using the broadening function $\Delta^{\infty}_{dis}$.
The stellar lines in this region were disentangled as a superposition of two
systems of lines using $\Delta^{1}_{dis}$   
with free radial velocities and each one with its own free line-strength
factors. To avoid uncertainties in low Fourier modes, which could cause
anticorrelated distortions of the stellar and telluric component
continua, we used disentangling constrained by a template (cf. Hadrava
2006b) for the telluric lines. The template had been calculated by
disentangling spectra of the star 68u~Her (i.e. spectroscopic
binary HD 156633, also taken with the Ond\v{r}ejov
2m-telescope), which gives a telluric spectrum with a satisfactorily flat
continuum. The differences between the prescribed annual motion of the
telluric lines and the radial velocities of these lines disentangled as
the best fits of individual exposures were then used as corrections of
the wavelength scale in the preparation of spectral regions for subsequent
disentangling of the stellar lines. Owing to application of a method
of enhanced precision (Hadrava 2009), these corrections were found with
sub-pixel resolution. These wavelength corrections can be applied
to exposures with sufficiently strong telluric lines only. In our case,
the depth of telluric lines in exposures where they are weakest is about
one half of their mean depth, so that the correction could be applied to
all exposures.
\begin{table}
 \caption[]{List of disentangled spectral lines}
  \label{tab2}
 $$
   \begin{array}{rclrccr}
     \hline
     \noalign{\smallskip}
 \lambda {\rm ~[\AA]~} & rvpb & {\rm multiplet} & \chi + \varepsilon {\rm ~[eV]~} & EW {\rm ~[\AA]~} &
  v_1 & s_1\hspace{2mm} \\
      \noalign{\smallskip}
      \hline
      \noalign{\smallskip}
6265.14 & 0.7 & {\rm Fe ~I~  (62)} &      2.18 & 0.107 & 19.89 & 0.402 \\ 
6343.71 & 0.8 & {\rm Ca ~I~  (53)} &      4.44 & 0.184 & 20.40 & 0.531 \\ 
6347.10 & 0.7 & {\rm Si ~II~  (2)} & 8.15+8.12 & 0.245 & 18.88 &-0.203 \\ 
6355.04 & 0.6 & {\rm Fe ~I~ (342)} &      2.83 & 0.110 & 19.82 & 0.508 \\ 
6358.69 & 0.6 & {\rm Fe ~I~  (13)} &      0.86 & 0.134 & 18.28 & 0.712 \\ 
6400.01 & 1.5 & {\rm Fe ~I~ (816)} &      3.60 & 0.263 & 21.42 & 0.319 \\ 
6456.39 & 0.8 & {\rm Fe ~II~ (74)} & 7.87+3.90 & 0.443 & 17.78 &-0.069 \\ 
6663.45 & 1.0 & {\rm Fe ~I~ (111)} &      2.42 & 0.188 & 20.67 & 0.384 \\ 
      \noalign{\smallskip}
      \hline
   \end{array}
     $$
\end{table}

 To disentangle the pulsational velocities in the observed spectra of
$\delta$~Cep we chose first several narrow spectral regions, each
one containing a single dominant spectral line listed in Table~\ref{tab2}.
These regions were sampled in 256 bins each with a step of radial velocity
per bin given in the Table ($rvpb$ in km/s). In some of the regions, blends
with some weak lines can be seen, which partly decrease the quality of
the disentangling, however, their influence can be neglected. The relatively
low spectral resolution of our original data (about 12 km/s) provides only
a rough sampling of line profiles in individual exposures, which have
half-depth widths comparable to the amplitude of the Doppler shifts, i.e.
only about 4 to 6 times larger.
The disentangled line profiles are significantly smoother due to
the averaging of a large number of the exposures. Nevertheless, the limited
quality of the data does not enable us to convincingly decide
which of the models $\Delta^{0}_{dis}$, $\Delta^{1}_{dis}$ and
$\Delta^{\infty}_{dis}$ best fits the observed line-profile variations
or even to search for their best linear combination, which could
be expected for more general limb-darkening within the line-profile.
We thus performed the disentangling of all regions for each of these
models separately. In all regions, the integrated (O--C)$^2$ of the
fitted spectra was the largest for the model $\Delta^{\infty}_{dis}$,
while the residual noise for the pulsational broadenings $\Delta^{0}_{dis}$
and $\Delta^{1}_{dis}$ was nearly the same (within about 4\%) without
any evident regular preference of one model or the other. We thus
illustrate the results on the case $\Delta^{1}_{dis}$, which
corresponds to the unit limb darkening and is thus the most advantageous
from the theoretical point of view (see above).

   \begin{figure}
   \centering
   \includegraphics[width=9cm]{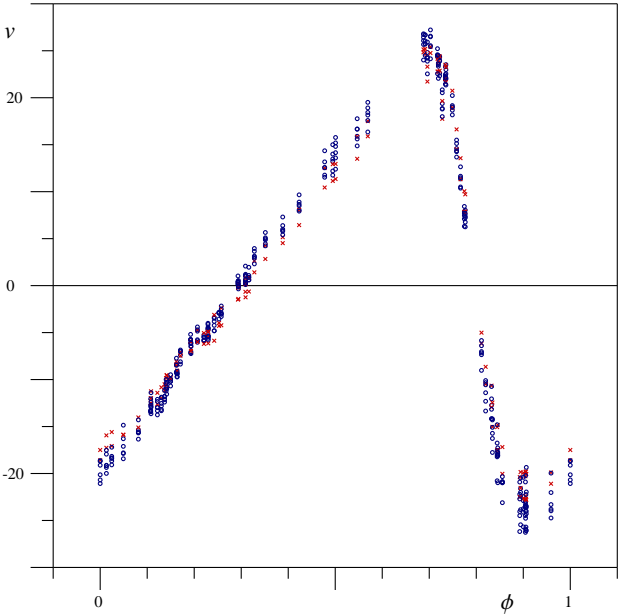}
   \setlength{\unitlength}{1mm}
   \put(-23,2){$\phi$}
   \put(-89,78){$v$}
   \caption{ Phase-dependence of pulsational velocity $v$ in km/s for
   individual spectral lines disentangled with $\Delta^{1}_{dis}$.
   The phase $\phi$ is labeled in cycles (i.e. the non-integer part
   of the ephemeris $E$). Lines of neutral atoms are marked by open circles
   (blue in electronic version), lines of ionized atoms by (red) crosses.
   }
 \label{obr2}
   \end{figure}

 Figure~\ref{obr2} shows the pulsational velocity disentangled
using the model $\Delta^{1}_{dis}$ which is given in the on-line
Table~\ref{tab3}. For the purpose of visualizing
the results (i.e. calculation of phase) we use the ephemeris
\begin{equation}
 HJD=2454697.0009+ 5.366341 \times E\; .
\end{equation}
Our data do not allow us to check or improve the period,
and the scatter in its published values (e.g. $P=5\fd{}3663159$ by
Moffett and Barnes, 1985, or $5\fd{}3662351$ according to the quadratic
ephemeris by Berdnikov and Ignatova, 2000)  
is not substantial for our present purpose. The reference
epoch is chosen to closely precede our observations and
coincide with the second exposure, which is relatively close to
the minimum radial velocity.

   \begin{figure}
   \centering
   \includegraphics[width=9cm]{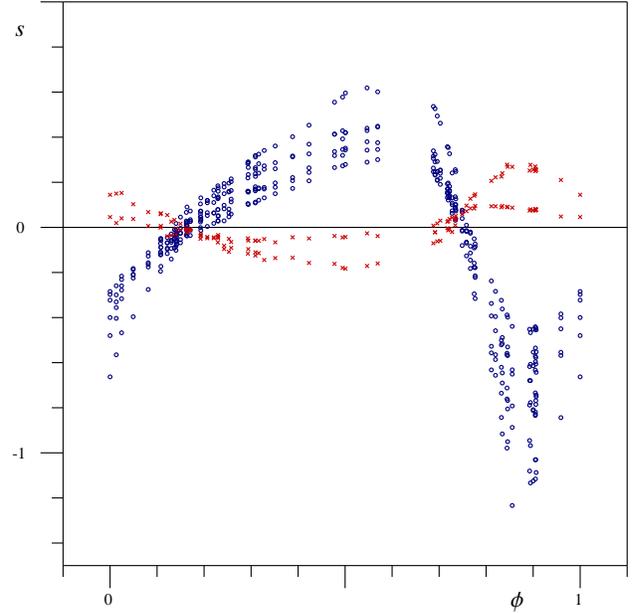}
   \setlength{\unitlength}{1mm}
   \put(-23,2){$\phi$}
   \put(-89,78){$s$}
   \caption{ Phase-dependence of line-strength factors $s$ for individual
   spectral lines disentangled with $\Delta^{1}_{dis}$. (The meaning
   of symbols is the same as in Fig.~\ref{obr2}.)
   }
 \label{obr3}
   \end{figure}
%

\onltab{3}{
\begin{table*}
 \caption[]{Pulsational velocities $v$ for individual lines (cf.
 Table~\ref{tab2}) and H$\alpha$ region at all exposures.}
  \label{tab3}
 $$
 \begin{array}{rrrrrrrrrrrrr}
 \hline
 \noalign{\smallskip}
 \begin{array}{c}{\rm HJD\hspace*{3mm}}\\ -2454000\end{array}
 & {\rm phase} & 6265.14 & 6343.71 & 6347.10 &
 6355.04 & 6358.69 & 6400.01 & 6456.39 & 6663.45 & H_{\alpha\,1} & H_{\alpha\,2} & H_{\alpha\,3} \\
      \noalign{\smallskip}
      \hline
      \noalign{\smallskip}
 698.3027 &.2426 &  -3.472 &  -4.805 &  -3.114 &  -4.231 &  -3.812 &  -3.377 &  -5.866 &  -4.813 & -10.724 &  -3.538 &  -3.581 \\
 702.3672 &.0000 & -19.120 & -20.131 & -18.591 & -18.690 & -18.623 & -21.070 & -17.497 & -20.683 & -20.926 & -18.053 & -21.937 \\
 708.4863 &.1403 & -10.826 & -10.841 &  -9.845 & -10.457 &  -9.998 & -10.685 & -10.559 & -11.224 & -16.223 &  -9.298 & -10.475 \\
 708.4925 &.1414 & -10.311 & -11.554 &  -9.525 & -11.006 & -10.281 &  -9.949 &  -9.568 & -10.841 & -16.093 &  -8.544 & -10.079 \\
 709.3067 &.2932 &    .463 &    .161 &    .131 &    .168 &    .355 &   1.027 &  -1.491 &    .367 &  -6.903 &   -.535 &    .282 \\
 709.3918 &.3090 &    .541 &   1.024 &    .803 &    .661 &   1.017 &   2.076 &   -.660 &   1.090 &  -5.914 &    .429 &   1.306 \\
 709.4934 &.3279 &   3.044 &   3.655 &   2.640 &   2.909 &   2.310 &   3.953 &   1.411 &   3.139 &  -3.749 &   2.340 &   3.406 \\
 710.2962 &.4775 &  11.764 &  13.170 &  12.519 &  12.540 &  11.533 &  14.355 &  10.445 &  12.580 &   6.654 &  10.167 &  13.655 \\
 710.3885 &.4947 &  12.200 &  14.001 &  12.923 &  13.476 &  11.755 &  15.001 &  11.141 &  13.225 &   7.582 &  10.522 &  14.431 \\
 713.3641 &.0492 & -16.299 & -18.433 & -15.865 & -17.112 & -14.856 & -17.865 & -15.884 & -17.800 & -21.821 & -14.746 & -17.823 \\
 719.2701 &.1498 &  -9.941 & -10.207 &  -9.739 &  -9.979 &  -9.482 & -10.167 &  -9.862 & -10.687 & -16.053 &  -9.057 &  -9.976 \\
 719.3365 &.1622 &  -8.472 &  -9.176 &  -8.075 &  -8.449 &  -7.727 &  -8.343 &  -8.468 &  -9.273 & -14.641 &  -7.795 &  -8.541 \\
 719.3823 &.1707 &  -8.104 &  -6.857 &  -7.089 &  -6.962 &  -7.039 &  -7.396 &  -7.464 &  -8.359 & -13.749 &  -7.148 &  -7.656 \\
 719.4983 &.1923 &  -6.140 &  -5.839 &  -5.920 &  -5.720 &  -5.739 &  -5.189 &  -5.965 &  -7.151 & -12.214 &  -5.771 &  -5.955 \\
 719.5025 &.1931 &  -7.041 &  -6.427 &  -6.718 &  -6.484 &  -6.072 &  -6.175 &  -6.978 &  -7.242 & -13.088 &  -6.391 &  -6.643 \\
 719.5764 &.2069 &  -5.763 &  -6.034 &  -4.739 &  -4.705 &  -4.399 &  -4.846 &  -6.002 &  -5.864 & -11.677 &  -5.091 &  -5.187 \\
 736.4506 &.3513 &   5.086 &   4.986 &   4.263 &   4.868 &   4.232 &   5.656 &   2.834 &   4.452 &  -2.272 &   4.177 &   5.114 \\
 737.2522 &.5007 &  14.810 &  14.151 &  12.956 &  13.590 &  12.394 &  15.191 &  11.347 &  15.769 &   9.054 &  11.995 &  15.898 \\
 737.4991 &.5467 &  15.897 &  16.675 &  15.857 &  15.609 &  14.873 &  17.766 &  13.501 &  16.647 &  11.422 &  13.820 &  18.197 \\
 737.6202 &.5693 &  18.153 &  18.430 &  17.510 &  17.579 &  16.354 &  19.509 &  15.891 &  18.914 &  13.497 &  15.659 &  20.182 \\
 738.2559 &.6877 &  25.667 &  26.431 &  25.142 &  26.082 &  24.013 &  26.733 &  24.830 &  26.803 &  21.757 &  23.688 &  27.721 \\
 738.3353 &.7025 &  25.542 &  26.555 &  24.756 &  25.440 &  24.145 &  26.485 &  25.407 &  27.217 &  23.200 &  24.194 &  28.507 \\
 738.4407 &.7222 &  22.621 &  24.298 &  22.876 &  23.889 &  22.326 &  24.393 &  24.408 &  23.966 &  23.626 &  22.297 &  27.316 \\
 738.5058 &.7343 &  21.891 &  22.284 &  22.006 &  22.177 &  21.452 &  22.448 &  23.506 &  23.296 &  24.665 &  20.887 &  26.895 \\
 738.5121 &.7355 &  22.000 &  22.983 &  21.673 &  22.090 &  21.350 &  22.549 &  23.257 &  23.523 &  24.795 &  20.794 &  26.921 \\
 738.5852 &.7491 &  18.959 &  19.164 &  18.943 &  19.028 &  18.171 &  18.670 &  20.746 &  20.229 &  24.993 &  18.405 &  24.877 \\
 744.3990 &.8325 & -12.053 & -12.730 & -12.948 & -13.166 & -10.740 & -14.169 & -10.648 & -14.182 &  15.896 & -11.724 & -16.115 \\
 751.2407 &.1074 & -13.394 & -13.437 & -12.515 & -12.752 & -12.051 & -13.271 & -11.996 & -13.347 & -17.913 & -11.563 & -13.046 \\
 751.2445 &.1081 & -13.639 & -12.864 & -11.994 & -12.394 & -11.397 & -12.946 & -11.251 & -12.639 & -17.725 & -11.017 & -12.618 \\
 752.2419 &.2940 &   -.351 &   -.150 &   -.213 &   -.023 &   -.118 &    .540 &  -1.416 &    .192 &  -7.474 &    .391 &    .310 \\
 752.3239 &.3093 &    .104 &    .517 &    .161 &    .235 &    .368 &   1.205 &  -1.246 &    .242 &  -6.694 &    .390 &    .755 \\
 752.3612 &.3162 &    .601 &    .873 &    .901 &    .883 &   1.079 &   1.952 &   -.628 &    .618 &  -5.854 &    .430 &   1.406 \\
 757.2615 &.2294 &  -4.640 &  -4.563 &  -4.879 &  -4.458 &  -4.136 &  -4.008 &  -5.363 &  -4.740 & -11.266 &  -4.572 &  -4.490 \\
 758.3017 &.4232 &   8.058 &   8.891 &   8.077 &   8.648 &   7.927 &   9.663 &   6.442 &   8.495 &   2.435 &   6.836 &   9.369 \\
 761.4677 &.0132 & -19.167 & -19.361 & -17.262 & -17.943 & -17.526 & -19.989 & -15.940 & -19.126 & -22.674 & -15.931 & -20.340 \\
 771.2950 &.8445 & -16.757 & -18.029 & -17.489 & -17.553 & -14.807 & -20.774 & -15.062 & -18.224 &  10.474 & -15.843 & -21.765 \\
 771.3017 &.8457 & -17.270 & -17.912 & -17.540 & -18.093 & -17.744 & -20.982 & -14.826 & -18.092 &   9.583 & -16.326 & -22.067 \\
 772.2613 &.0245 & -18.332 & -18.172 & -17.079 & -18.665 & -17.204 & -19.081 & -15.587 & -18.398 & -21.602 & -15.372 & -19.205 \\
 774.2135 &.3883 &   6.001 &   5.869 &   5.153 &   5.797 &   5.404 &   7.304 &   4.523 &   6.392 &   -.147 &   5.395 &   7.202 \\
 776.1982 &.7582 &  14.591 &  15.493 &  14.505 &  14.258 &  14.348 &  13.684 &  16.629 &  15.125 &  23.436 &  14.231 &  21.215 \\
 776.2430 &.7665 &  11.360 &  11.644 &  11.325 &  11.487 &  10.499 &  10.375 &  13.564 &  12.644 &  23.432 &  11.784 &  18.674 \\
 776.2868 &.7747 &   7.889 &   7.464 &   7.835 &   7.097 &   7.709 &   6.312 &  10.046 &   8.428 &  22.699 &   8.657 &  14.885 \\
 776.2972 &.7766 &   7.360 &   6.757 &   7.886 &   7.655 &   7.222 &   6.257 &   9.713 &   8.030 &  23.566 &   8.493 &  14.584 \\
 781.2311 &.6960 &  24.209 &  24.341 &  21.721 &  23.887 &  22.540 &  24.820 &  23.302 &  25.921 &  21.316 &  22.742 &  26.844 \\
 798.3846 &.8925 & -22.209 & -20.912 & -22.504 & -22.707 & -24.052 & -26.179 & -20.384 & -24.487 &  -4.586 & -21.204 & -28.487 \\
 798.3939 &.8943 & -22.432 & -21.594 & -21.541 & -21.999 & -20.465 & -25.455 & -19.844 & -23.852 &  -4.747 & -20.621 & -27.909 \\
 798.4495 &.9046 & -22.934 & -23.639 & -22.419 & -23.410 & -20.900 & -26.265 & -20.292 & -24.475 &  -7.728 & -21.228 & -28.217 \\
 824.3317 &.7277 &  18.792 &  20.530 &  17.753 &  18.852 &  17.995 &  19.383 &  19.651 &  20.842 &  20.695 &  16.979 &  23.366 \\
 843.2488 &.2528 &  -2.825 &  -2.918 &  -3.969 &  -3.555 &  -2.882 &  -2.933 &  -4.288 &  -2.926 & -10.146 &  -2.626 &  -2.791 \\
 857.2159 &.8555 & -20.871 & -20.889 & -20.036 & -20.335 & -20.987 & -23.105 & -17.185 & -20.980 &   5.365 & -19.282 & -25.892 \\
 910.6409 &.8111 &  -7.020 &  -7.160 &  -6.173 &  -6.396 &  -5.840 &  -9.023 &  -5.008 &  -7.355 &  20.318 &  -5.737 &  -6.244 \\
 923.5714 &.2207 &  -5.783 &  -5.871 &  -5.055 &  -5.467 &  -5.484 &  -5.258 &  -6.204 &  -5.840 & -12.770 &  -5.549 &  -5.808 \\
 923.6211 &.2299 &  -5.610 &  -5.789 &  -4.929 &  -5.372 &  -5.455 &  -4.547 &  -6.149 &  -5.344 & -12.815 &  -5.218 &  -5.527 \\
 927.5339 &.9591 & -23.305 & -22.027 & -21.087 & -23.797 & -19.956 & -24.735 & -19.891 & -23.971 & -21.156 & -19.528 & -25.434 \\
 928.4066 &.1217 & -12.945 & -13.761 & -11.411 & -12.335 & -12.105 & -13.326 & -12.596 & -12.991 & -18.469 & -10.608 & -12.469 \\
 928.4523 &.1302 & -12.809 & -13.317 & -10.827 & -11.937 & -11.876 & -12.539 & -11.972 & -13.189 & -18.086 & -10.433 & -12.017 \\
 928.4885 &.1370 & -12.083 & -12.143 & -10.498 & -11.798 & -11.156 & -11.789 & -11.150 & -12.400 & -17.737 & -10.201 & -11.629 \\
 928.6318 &.1637 &  -9.680 &  -9.726 &  -9.079 &  -9.399 &  -9.103 &  -9.293 &  -9.046 &  -9.384 & -14.967 &  -8.243 &  -8.868 \\
 931.4636 &.6914 &  25.727 &  25.717 &  24.632 &  26.043 &  24.502 &  26.672 &  25.196 &  26.687 &  21.251 &  23.321 &  27.381 \\
 931.6042 &.7176 &  24.333 &  24.484 &  22.787 &  24.512 &  22.557 &  24.579 &  24.155 &  25.221 &  23.197 &  22.469 &  27.464 \\
 931.6139 &.7194 &  23.349 &  23.641 &  22.779 &  23.520 &  22.059 &  23.678 &  23.775 &  24.484 &  23.100 &  21.896 &  27.099 \\
 932.5844 &.9002 & -22.441 & -22.897 & -22.678 & -23.743 & -20.409 & -25.682 & -19.925 & -24.120 &  -7.148 & -20.443 & -27.991 \\
 932.6097 &.9049 & -23.106 & -24.767 & -22.705 & -23.376 & -20.218 & -25.918 & -19.832 & -24.021 &  -8.526 & -20.536 & -27.756 \\
 932.6138 &.9057 & -23.617 & -25.700 & -22.690 & -23.541 & -22.236 & -26.092 & -19.773 & -23.632 &  -8.680 & -20.248 & -27.686 \\
 932.6178 &.9064 & -21.966 & -24.284 & -22.814 & -23.471 & -19.347 & -26.057 & -20.040 & -24.018 &  -8.634 & -20.353 & -27.959 \\
 933.5557 &.0812 & -15.562 & -15.682 & -15.102 & -15.644 & -14.291 & -16.345 & -14.030 & -15.348 & -20.031 & -12.778 & -15.282 \\
 934.5024 &.2576 &  -3.043 &  -3.294 &  -2.396 &  -2.486 &  -3.021 &  -2.173 &  -4.223 &  -2.895 & -10.500 &  -2.929 &  -2.991 \\
 937.5199 &.8199 & -10.328 & -12.313 & -10.500 & -10.560 & -10.105 & -13.348 &  -8.644 & -11.404 &  16.376 &  -9.508 & -11.648 \\
 937.5973 &.8344 & -14.296 & -15.723 & -15.037 & -15.085 & -14.187 & -17.760 & -12.443 & -16.271 &  12.039 & -13.840 & -18.320 \\
      \noalign{\smallskip}
      \hline
   \end{array}
     $$
\end{table*}
}

 The fit of the observed spectra by the standard disentangling
($\Delta^{\infty}_{dis}$) is invariant with respect to adding a constant
to all disentangled radial velocities and a simultaneous shift of the
disentangled spectrum in the logarithmic wavelength for the corresponding
value. If the position of the stellar spectrum disentangled from all
exposures (transformed first into heliocentric wavelength scale) is fixed
to the laboratory wavelengths of identified stellar lines, the measured
radial velocities correspond to instantaneous heliocentric radial
velocities. Neglecting the systematic errors caused by the line-profile
distortions, the heliocentric radial velocity of the star could be
estimated as the velocity averaged over the pulsational period
\begin{equation}
 \bar{v}\equiv\frac{1}{P}\int_{0}^{P}v(t)dt\; .
\end{equation}
Having the values of velocities measured with some errors in several
discrete times only, we fit them using least-squares method in the
standard manner by Fourier series
(cf. Schaltenbrand and Tammann 1971, or Moffett and Barnes 1985, etc.)
\begin{equation}\label{Fourser}
 v(t)\simeq v_{0}+\sum_{k=1}^{K}v_{k}\sin(2\pi kt/P+\phi_{k})\; .
\end{equation}
The mean velocity $\bar{v}$ is then given by the first term $v_{0}$.
We have chosen $K=5$, because the residual $(O-C)^{2}$ of the fit
decreased significantly with each additional harmonic term up to this
value, but its value remains practically constant (given by the rms.
error of the measurements) for higher degree of the expansion.
We also used the same Fourier series for the line-strength
factors.

 Unlike the standard disentangling, the disentangling with broadening
functions $\Delta^{0}_{dis}$ and $\Delta^{1}_{dis}$ is sensitive to
the value of the true radial velocity of the star's centre of mass.
At the phases when the pulsational velocity is equal to zero, the consequent
line-profile distortion disappears, the instantaneous line-profile coincides
(within the errors) with the disentangled one and its Doppler shift is given
by the overall motion of the star only. The Doppler shift of the disentangled
spectrum should thus correspond to the intrinsic radial velocity of the
star. However, because a systematic shift of all pulsational velocities
can be relatively well compensated by a shift of the disentangled spectrum,
the convergence of the solution to the true radial velocity is very slow
and the value of the velocity is poorly defined, in particular for the rough
sampling of the line profiles in our spectra. (To speed up the convergence
we introduced an additive term in the velocity, which is possible to converge
explicitly by the simplex method.) The values for which we
achieve the best fit of the observed spectra differ somewhat from
the values $v_{0}$ found from the fit by Eq.~(\ref{Fourser}). This
discrepancy may be due to the mentioned observational errors, but it may
also reflect an influence of effects neglected in the simple models given
by the broadening functions $\Delta^{0}_{dis}$ and $\Delta^{1}_{dis}$.
For instance the gradients of pulsational velocities or the stellar wind
overimposed on the pulsations may contribute a 
distortion resembling P-Cyg shape to the line profile. 
More precise measurements as well as models
of line formation will be needed to solve this question.

 In Table~\ref{tab2} we give the amplitude $v_1$ of the first periodic term
in the expansion given by Eq.~(\ref{Fourser}). It can be seen that this
amplitude is smaller for the lines of ionized elements and it is slightly
correlated with the equivalent width of the lines of neutral elements.
This result indicates that the amplitude of pulsational velocities depends
on the depth in the atmosphere where the line is formed (cf. Butler et al.
1996, Petterson et al. 2005, and references therein). This also
qualitatively agrees with the correlation between line-depths and
amplitudes of radial-velocity curves found in Cepheids by Nardetto et
al. (2007 and 2009).

\onltab{4}{
\begin{table*}
 \caption[]{Line-strength factors for individual lines (cf.
 Table~\ref{tab2}) and H$\alpha$ region at all exposures.}
  \label{tab4}
 $$
   \begin{array}{rrrrrrrrrrrrr}
     \hline
     \noalign{\smallskip}
 \begin{array}{c}{\rm HJD\hspace*{3mm}}\\ -2454000\end{array}
 & {\rm phase} & 6265.14 & 6343.71 & 6347.10 &
 6355.04 & 6358.69 & 6400.01 & 6456.39 & 6663.45 & H_{\alpha\,1} & H_{\alpha\,2} & H_{\alpha\,3} \\
      \noalign{\smallskip}
      \hline
      \noalign{\smallskip}
 698.3027 &.2426 &  .0822 &  .1945 & -.0958 &  .1495 &  .1788 &  .0290 & -.0812 &  .0981 &  .1646 & -.0497 & -.0258 \\
 702.3672 &.0000 & -.3228 & -.3998 &  .1446 & -.4800 & -.6630 & -.2845 &  .0459 & -.2972 & -.5038 &  .1219 &  .0517 \\
 708.4863 &.1403 & -.0631 &  .0173 & -.0047 & -.0120 & -.1017 & -.0843 & -.0182 & -.0507 & -.0129 &  .0160 &  .0057 \\
 708.4925 &.1414 & -.0621 & -.0422 & -.0268 & -.0314 & -.0726 & -.0748 & -.0320 & -.0200 & -.0290 &  .0138 &  .0024 \\
 709.3067 &.2932 &  .1607 &  .2827 & -.1163 &  .2193 &  .2876 &  .1056 & -.0488 &  .1595 &  .2507 & -.0651 & -.0263 \\
 709.3918 &.3090 &  .1620 &  .2648 & -.1247 &  .2376 &  .3187 &  .1084 & -.0597 &  .1801 &  .2522 & -.0712 & -.0312 \\
 709.4934 &.3279 &  .1702 &  .2728 & -.1448 &  .2580 &  .3411 &  .1310 & -.0519 &  .1772 &  .2757 & -.0802 & -.0356 \\
 710.2962 &.4775 &  .3363 &  .4138 & -.1619 &  .4130 &  .5553 &  .2653 & -.0381 &  .3183 &  .4408 & -.1235 & -.0453 \\
 710.3885 &.4947 &  .3457 &  .4286 & -.1795 &  .3925 &  .5778 &  .2728 & -.0447 &  .3195 &  .4484 & -.1263 & -.0475 \\
 713.3641 &.0492 & -.2249 & -.2079 &  .1028 & -.2110 & -.3964 & -.1825 &  .0373 & -.1871 & -.2661 &  .0813 &  .0378 \\
 719.2701 &.1498 & -.0248 &  .0455 &  .0164 & -.0269 & -.0686 & -.0543 & -.0114 & -.0041 &  .0258 &  .0100 &  .0055 \\
 719.3365 &.1622 & -.0085 &  .0665 & -.0107 &  .0285 &  .0198 & -.0346 & -.0035 &  .0099 &  .0585 &  .0068 &  .0061 \\
 719.3823 &.1707 & -.0118 &  .0725 & -.0072 &  .0273 &  .0343 & -.0359 & -.0136 &  .0258 &  .0646 & -.0012 & -.0001 \\
 719.4983 &.1923 &  .0241 &  .1042 & -.0507 &  .0600 &  .0720 & -.0009 & -.0390 &  .0466 &  .1049 & -.0176 & -.0089 \\
 719.5025 &.1931 &  .0314 &  .1309 & -.0362 &  .0719 &  .0455 & -.0135 & -.0307 &  .0359 &  .1078 & -.0157 & -.0061 \\
 719.5764 &.2069 &  .0527 &  .1196 & -.0483 &  .1092 &  .1192 &  .0041 & -.0419 &  .0652 &  .1261 & -.0264 & -.0130 \\
 736.4506 &.3513 &  .1779 &  .2870 & -.1361 &  .2601 &  .3609 &  .1495 & -.0662 &  .1972 &  .2877 & -.0917 & -.0421 \\
 737.2522 &.5007 &  .3512 &  .4208 & -.1832 &  .4199 &  .5960 &  .2792 & -.0423 &  .3434 &  .4505 & -.1257 & -.0469 \\
 737.4991 &.5467 &  .3798 &  .4295 & -.1712 &  .4397 &  .6185 &  .2893 & -.0264 &  .3426 &  .4776 & -.1349 & -.0505 \\
 737.6202 &.5693 &  .3756 &  .4485 & -.1599 &  .4454 &  .6012 &  .3005 & -.0385 &  .3456 &  .4745 & -.1365 & -.0540 \\
 738.2559 &.6877 &  .3053 &  .3395 & -.0703 &  .3159 &  .5363 &  .2469 &  .0079 &  .2631 &  .4079 & -.0913 & -.0403 \\
 738.3353 &.7025 &  .2544 &  .2399 & -.0595 &  .2531 &  .4621 &  .1909 &  .0302 &  .2182 &  .3824 & -.0810 & -.0408 \\
 738.4407 &.7222 &  .1818 &  .1591 & -.0076 &  .1638 &  .3267 &  .1410 &  .0368 &  .1314 &  .3433 & -.0588 & -.0372 \\
 738.5058 &.7343 &  .1359 &  .0316 &  .0093 &  .0895 &  .2559 &  .1008 &  .0392 &  .0850 &  .3233 & -.0448 & -.0369 \\
 738.5121 &.7355 &  .1265 &  .0375 &  .0330 &  .0729 &  .2478 &  .1011 &  .0511 &  .0796 &  .3302 & -.0435 & -.0353 \\
 738.5852 &.7491 &  .0666 & -.0804 &  .0751 & -.0164 &  .0764 &  .0393 &  .0582 & -.0128 &  .2784 & -.0145 & -.0283 \\
 744.3990 &.8325 & -.5208 & -.7354 &  .2252 & -.6206 & -.8432 & -.3233 &  .0891 & -.5007 & -.5875 &  .2436 &  .1080 \\
 751.2407 &.1074 & -.1120 & -.0517 &  .0590 & -.1096 & -.1948 & -.1219 & -.0007 & -.0718 & -.0941 &  .0380 &  .0145 \\
 751.2445 &.1081 & -.0881 & -.0494 &  .0627 & -.0886 & -.1729 & -.1202 &  .0065 & -.0581 & -.0913 &  .0377 &  .0150 \\
 752.2419 &.2940 &  .1563 &  .2657 & -.0964 &  .2344 &  .2906 &  .1015 & -.0583 &  .1596 &  .2276 & -.0672 & -.0287 \\
 752.3239 &.3093 &  .1694 &  .2572 & -.1106 &  .2262 &  .3124 &  .1120 & -.0581 &  .1696 &  .2443 & -.0740 & -.0329 \\
 752.3612 &.3162 &  .1713 &  .2825 & -.1142 &  .2405 &  .3249 &  .1205 & -.0593 &  .1746 &  .2609 & -.0765 & -.0336 \\
 757.2615 &.2294 &  .0788 &  .1843 & -.0666 &  .1306 &  .1272 &  .0300 & -.0417 &  .0791 &  .1495 & -.0353 & -.0166 \\
 758.3017 &.4232 &  .2640 &  .3692 & -.1577 &  .3261 &  .4534 &  .2066 & -.0491 &  .2440 &  .3620 & -.1138 & -.0507 \\
 761.4677 &.0132 & -.3306 & -.4034 &  .1500 & -.3557 & -.5652 & -.2592 &  .0193 & -.2993 & -.4691 &  .1000 &  .0386 \\
 771.2950 &.8445 & -.5584 & -.7611 &  .2783 & -.7125 & -.9788 & -.3885 &  .0890 & -.5658 & -.8165 &  .2429 &  .1093 \\
 771.3017 &.8457 & -.5683 & -.8052 &  .2690 & -.7677 & -.9505 & -.3981 &  .0913 & -.5304 & -.8146 &  .2429 &  .1100 \\
 772.2613 &.0245 & -.2594 & -.2781 &  .1527 & -.3165 & -.4674 & -.2314 &  .0402 & -.2163 & -.4133 &  .1053 &  .0466 \\
 774.2135 &.3883 &  .2071 &  .3127 & -.1345 &  .3166 &  .4022 &  .1728 & -.0439 &  .2210 &  .3391 & -.1006 & -.0412 \\
 776.1982 &.7582 & -.0143 & -.1265 &  .0966 & -.0651 & -.0183 &  .0098 &  .0719 & -.0697 &  .2625 &  .0126 & -.0146 \\
 776.2430 &.7665 & -.0260 & -.1824 &  .1273 & -.1452 & -.0833 & -.0207 &  .0843 & -.1136 &  .2290 &  .0452 &  .0006 \\
 776.2868 &.7747 & -.0914 & -.2949 &  .1402 & -.1773 & -.1585 & -.0507 &  .0838 & -.1767 &  .1790 &  .0836 &  .0207 \\
 776.2972 &.7766 & -.0859 & -.3156 &  .1466 & -.2221 & -.2148 & -.0736 &  .0954 & -.1786 &  .1704 &  .0951 &  .0262 \\
 781.2311 &.6960 &  .2466 &  .2483 & -.0620 &  .2925 &  .4941 &  .2083 &  .0174 &  .2111 &  .3797 & -.0863 & -.0433 \\
 798.3846 &.8925 & -.6794 & -.9454 &  .2519 & -.7883 &-1.0806 & -.4518 &  .0747 & -.6174 &-1.1268 &  .2184 &  .0990 \\
 798.3939 &.8943 & -.6798 & -.9679 &  .2774 & -.7770 &-1.1335 & -.4669 &  .0799 & -.6073 &-1.1220 &  .2183 &  .0995 \\
 798.4495 &.9046 & -.6331 & -.8344 &  .2701 & -.7342 &-1.0309 & -.4408 &  .0748 & -.5710 &-1.0990 &  .2093 &  .0942 \\
 824.3317 &.7277 &  .1312 &  .0625 & -.0206 &  .1002 &  .2414 &  .1081 &  .0435 &  .0733 &  .3419 & -.0566 & -.0400 \\
 843.2488 &.2528 &  .0637 &  .1961 & -.1087 &  .1569 &  .1600 &  .0467 & -.0507 &  .0811 &  .1681 & -.0553 & -.0287 \\
 857.2159 &.8555 & -.6509 & -.8872 &  .2686 & -.7922 &-1.2336 & -.4391 &  .0873 & -.6319 & -.9423 &  .2389 &  .1094 \\
 910.6409 &.8111 & -.3374 & -.5566 &  .2064 & -.5274 & -.6323 & -.2376 &  .0936 & -.3928 & -.2223 &  .2320 &  .1034 \\
 923.5714 &.2207 &  .0640 &  .1649 & -.0442 &  .1104 &  .1402 &  .0178 & -.0479 &  .0640 &  .1309 & -.0367 & -.0191 \\
 923.6211 &.2299 &  .0494 &  .1752 & -.0479 &  .1604 &  .1351 &  .0278 & -.0341 &  .0519 &  .1435 & -.0438 & -.0223 \\
 927.5339 &.9591 & -.4009 & -.5680 &  .2103 & -.5540 & -.8438 & -.3840 &  .0482 & -.4500 & -.8135 &  .1525 &  .0638 \\
 928.4066 &.1217 & -.0681 & -.0455 &  .0547 & -.0924 & -.1163 & -.0973 & -.0391 & -.0687 & -.0858 &  .0124 & -.0046 \\
 928.4523 &.1302 & -.0665 & -.0292 &  .0253 & -.0592 & -.1405 & -.0934 & -.0359 & -.0576 & -.0593 &  .0078 & -.0055 \\
 928.4885 &.1370 & -.0520 & -.0104 &  .0332 & -.0482 & -.0913 & -.0898 & -.0403 & -.0325 & -.0387 &  .0029 & -.0076 \\
 928.6318 &.1637 & -.0262 &  .0903 &  .0027 &  .0230 &  .0079 & -.0384 & -.0236 &  .0081 &  .0404 & -.0035 & -.0049 \\
 931.4636 &.6914 &  .2908 &  .2347 & -.0218 &  .3235 &  .5266 &  .2320 & -.0214 &  .2465 &  .3901 & -.1184 & -.0646 \\
 931.6042 &.7176 &  .1931 &  .1321 & -.0038 &  .1705 &  .3577 &  .1551 &  .0246 &  .1518 &  .3506 & -.0804 & -.0511 \\
 931.6139 &.7194 &  .1966 &  .1212 & -.0128 &  .1721 &  .3544 &  .1502 &  .0286 &  .1391 &  .3499 & -.0775 & -.0501 \\
 932.5844 &.9002 & -.6140 & -.8125 &  .2560 & -.8086 &-1.1247 & -.4521 &  .0753 & -.5533 &-1.1375 &  .2081 &  .0925 \\
 932.6097 &.9049 & -.5955 & -.8288 &  .2492 & -.8191 &-1.1159 & -.4518 &  .0821 & -.5350 &-1.0868 &  .2048 &  .0914 \\
 932.6138 &.9057 & -.6062 & -.7851 &  .2597 & -.7683 &-1.0302 & -.4456 &  .0788 & -.5772 &-1.1177 &  .2054 &  .0916 \\
 932.6178 &.9064 & -.6428 & -.7434 &  .2541 & -.7506 &-1.0873 & -.4546 &  .0782 & -.5523 &-1.1196 &  .2050 &  .0915 \\
 933.5557 &.0812 & -.1641 & -.1123 &  .0679 & -.1756 & -.2754 & -.1572 &  .0065 & -.1275 & -.1884 &  .0491 &  .0165 \\
 934.5024 &.2576 &  .1131 &  .2151 & -.0909 &  .1636 &  .2027 &  .0592 & -.0651 &  .1253 &  .1865 & -.0594 & -.0291 \\
 937.5199 &.8199 & -.3956 & -.5863 &  .2219 & -.5648 & -.6562 & -.2838 &  .0934 & -.4585 & -.3727 &  .2369 &  .1075 \\
 937.5973 &.8344 & -.5158 & -.6904 &  .2402 & -.6457 & -.9161 & -.3619 &  .0952 & -.4885 & -.6210 &  .2414 &  .1099 \\
      \noalign{\smallskip}
      \hline
   \end{array}
     $$
\end{table*}

}

 Similarly, the line-strength factors (see Fig.~\ref{obr3} and
on-line Table~\ref{tab4}) of the lines were expanded into
the Fourier series and the amplitudes $s_1$ of the first component
are also given in Table~\ref{tab2}. The negative values of this amplitude
are assigned to the lines which are approximately in an antiphase
with the other lines. These are the two lines of ionized elements
in our set, for which $\phi_1\simeq -35^{\circ}$, while for the lines
of neutral atoms we have $\phi_1\simeq 165^{\circ}$. This amplitude
(anti-)correlates well with the excitation potential
$\varepsilon$ (+ ionization potential $\chi$, cf. Moore et al. 1966) of
the lower level of the line. The largest deviation can be seen for
the Ca~I line, which, however, may be blended with the Fe~I~(169) line
6344.15~{\AA} with $\varepsilon=2.42$~eV. These variations of line
strengths measured by the method of relative line photometry (Hadrava
1997) thus enable one to easily find the changes of
atmosphere temperature in the course of the Cepheid pulsation period
(cf. Krockenberger et al., 1998, Kovtyukh and Gorlova 2000).

   \begin{figure}
   \centering
   \includegraphics[width=9cm]{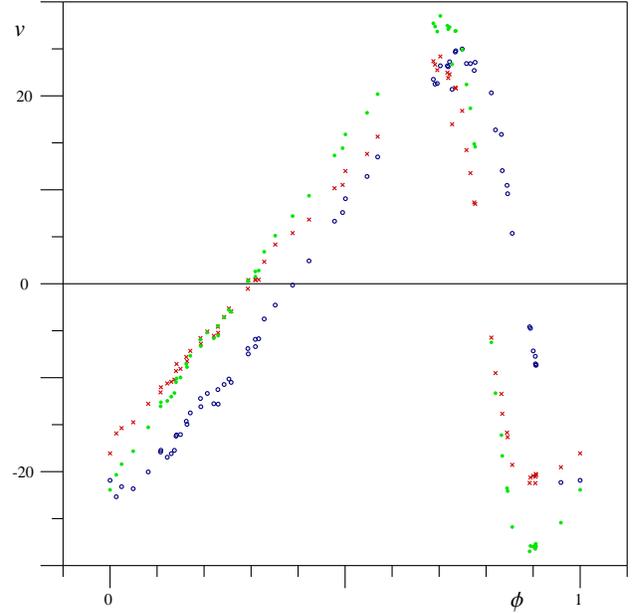}
   \setlength{\unitlength}{1mm}
   \put(-23,2){$\phi$}
   \put(-89,78){$v$}
   \caption{ Phase-dependence of pulsational velocity in km/s for
   spectral lines in the H$\alpha$ region. The one-component solution
   is marked by (green) full circles, low- and high- excitation
   components of the two-component disentangling are marked by
   (blue) open circles and (red) crosses, resp.
   }
 \label{obr4}
   \end{figure}
%

   \begin{figure}
   \centering
   \includegraphics[width=9cm]{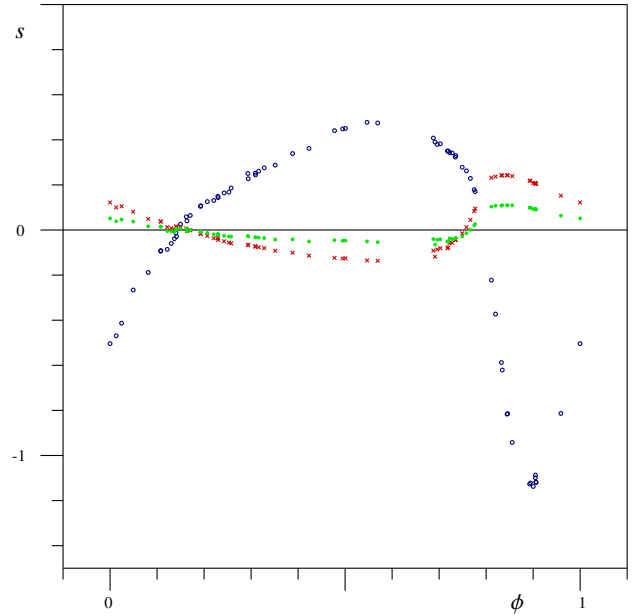}
   \setlength{\unitlength}{1mm}
   \put(-23,2){$\phi$}
   \put(-89,78){$s$}
   \caption{ Phase-dependence of line-strength factors $s$ for
   spectral lines in the H$\alpha$ region. (The meaning of symbols
   is the same as in Fig.~\ref{obr4}.)
   }
 \label{obr5}
   \end{figure}

\begin{figure}
   \centering
   \includegraphics[width=9cm]{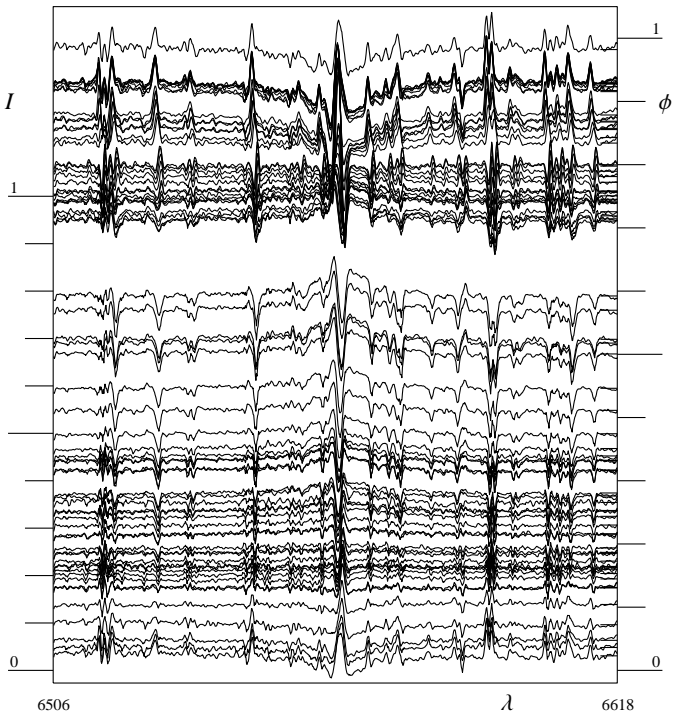}
   \setlength{\unitlength}{1mm}
   \put(-23,0){$\lambda$}
   \put(-89,80){$I$}
   \put(-2,80){$\phi$}
   \caption{ Residual spectra in the H$\alpha$ region after
   1-component (+ telluric spectrum) disentangling. The vertical
   offset is proportional to the phase.
   }
 \label{obr6}
   \end{figure}

 As a next step we performed disentangling of a wide spectral
region (6506--6618~{\AA}) around the H$\alpha$ line. We chose
sampling in 1024 bins with a step of 5km/s per bin. The telluric
lines were disentangled using a template obtained from spectra of
68u~Her again. The pulsational velocities and line-strength
factors obtained by disentangling of the whole spectral region into
one stellar component are given in Tables~\ref{tab3} and \ref{tab4}
(in columns labeled H$_{\alpha\, 3}$) and drawn (by full circles
in green) in Fig.~\ref{obr4}
and \ref{obr5}, respectively. The semiamplitude of the
first Fourier mode of the pulsational velocity is $v_{1}=21.40$~km/s;
the line-strength factors have a semiamplitude $s_{1}=0.062$ and phase
shift ($\phi_{1}\simeq -20^{\circ}$) comparable to that of the ionized
atoms in Fig.~\ref{obr3}. The phase dependence of residual spectra
is drawn in Fig.~\ref{obr6}. (The wavelength $\lambda$ is drawn
on a logarithmic scale and labeled in \AA.) It can be seen here
that the strength of the H$\alpha$ wide wings is almost in antiphase with
its narrow core as well as with strengths of weaker metallic lines.
The residuals in the wings are thus in absorption and the narrow
lines in emission around phase 0.0 around maximum expansion (cf.
the bottom and uppermost lines) and it is reversed around phase
0.5 at the infall (the lines around the middle of the Figure).
Residuals of some of the narrow lines have a shape of P-Cyg or
inverse P-Cyg profiles in some phases. This indicates that the Doppler
shifts of these lines differ from the other lines.

\begin{figure}
   \centering
   \includegraphics[width=9cm]{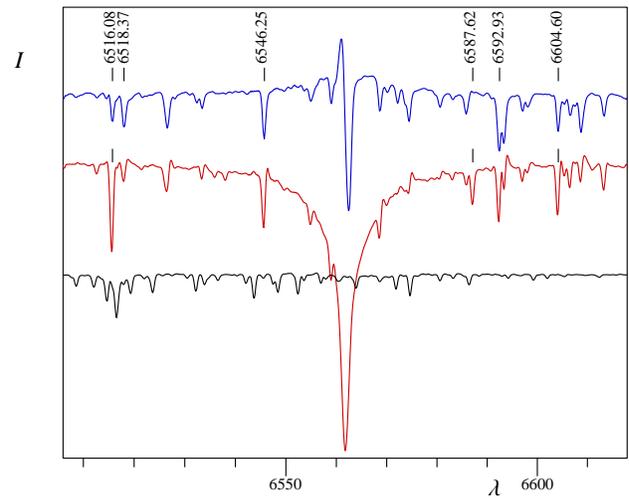}
   \setlength{\unitlength}{1mm}
   \put(-26,1){$\lambda$}
   \put(-89,58){$I$}
   \caption{ Spectrum of $\delta$~Cep in the H$_\alpha$ region disentangled
   into two components. The top (blue) line corresponds to the
   low-excitation component, the middle (red) line to the high-excitation
   component and the bottom (black) line to the telluric component.
   }
 \label{obr7}
   \end{figure}

 To decrease the residual spectrum, we must allow each line to vary
its line strengths in agreement with the phase-locked changes of the
atmosphere temperature. We thus also disentangled the H$\alpha$
region into two independent components, starting with the two solutions
from Table~\ref{tab2}, which are the opposite extremes in $s_1$,
i.e. the solution for the line Fe~I~(13) 6358.69~{\AA} and Si~II~(2)
6347.10~{\AA}. The spectrum splits into two components, as it is
shown by the upper two lines in Fig.~\ref{obr7}. The third very bottom
line in this Figure gives the telluric spectrum. It can be seen here
that the H$\alpha$ line is contained mostly in the second component
obtained from the initial approximation corresponding to the Si~II line,
while the narrow metallic lines are distributed between both components,
each one to a different proportion. A small contribution to the core
of H$\alpha$ also appears in the first spectrum. This contribution has
an inverse P-Cyg shape.

 We let the velocities and line-strengths of both these components in
all spectra converge to the best fit. The (O--C)$^{2}$ of residual
spectra decreased to 12\% in comparison with the previous one-component
(+ 1 telluric component) solution. The resulting velocities are
labeled as H$_{\alpha\,1}$ or H$_{\alpha\,2}$ in Table~\ref{tab3} and
drawn in Fig.~\ref{obr4} by (blue) open circles or (red) crosses for
the components developed from the solutions for neutral and ionized atoms,
respectively. The first Fourier modes of the pulsational velocities are
$v_{1}=20.18$~km/s or $v_{1}=17.38$~km/s, resp.
The phase shift of the first mode of velocities converged to
$\phi_{1}\simeq 228^{\circ}$ for the first component, unlike the values
$\phi_{1}\simeq 197^{\circ}\pm 3^{\circ}$ found for the second component
of H$\alpha$ region, its one-component solution, as well as for the
solutions of individual lines listed in Table~\ref{tab2} (with the
exception of the value $\simeq 203^{\circ}$ for the line Fe~II).
The line-strength factors (given in Table~\ref{tab4} and
Fig.~\ref{obr5}) have
semiamplitudes $s_{1}=0.577$ with phase shift ($\phi_{1}=166^{\circ}$)
comparable to that of neutral atoms in Fig.~\ref{obr3} for the first
component and $s_{1}=-0.152$ for the second component. The negative sign
here denotes again the approximate antiphase ($\phi_{1}\simeq-20^{\circ}$,
cf. Fig.~\ref{obr5}). Let us note that the line C~I~(22) entirely appears
and the lines Fe~II~(40) and Sc~II~(19) are more pronounced in the second
component also containing the major part of the high excitation H$\alpha$
line ($\varepsilon=10.20$~eV), while the other low excitation Fe~I lines
identified in Fig.~\ref{obr7} and Table~\ref{tab5} (where estimates of
equivalent widths in the first and second component are given) are by
greater or equal part contained in the first ``low excitation" component
of the spectrum.

\begin{table}
 \caption[]{List of spectral lines disentangled in the H$\alpha$ region}
  \label{tab5}
 $$
   \begin{array}{rlrcc}
     \hline
     \noalign{\smallskip}
 \lambda {\rm ~[\AA]~} & {\rm multiplet} & \chi + \varepsilon {\rm ~[eV]~}
 & EW_1 {\rm ~[\AA]~} &  EW_2 {\rm ~[\AA]~}
 \\
      \noalign{\smallskip}
      \hline
      \noalign{\smallskip}
6516.08 & {\rm Fe~II~(40)} & 7.87+2.89 & 0.046 & 0.102 \\
6518.37 & {\rm Fe~I~(342)} &      2.83 & 0.069 & 0.020 \\
6546.25 & {\rm Fe~I~(268)} &      2.76 & 0.087 & 0.063 \\
6587.63 & {\rm C~I~(22) }  &      8.58 & -.004 & 0.033 \\
6592.93 & {\rm Fe~I~(268)} &      2.73 & 0.097 & 0.082 \\
6604.60 & {\rm Sc~II~(19)} & 6.54+1.36 & 0.058 & 0.064 \\
      \noalign{\smallskip}
      \hline
   \end{array}
     $$
\end{table}

 These results are consistent with those for individual lines presented
above and with the variability of radial velocities and equivalent widths
found using classical methods by different authors cited there.

\section{Conclusions}\label{Concl}

Our measurements of pulsational velocities of $\delta$~Cep using
a newly generalized version KOREL09 of the code for Fourier
disentangling and line-strength photometry confirmed that the phase
variations of velocities and strengths of different spectral lines
depend on the depth of their formation in the stellar atmosphere and
on the excitation potential of their lower levels.
This means that any elaboration of the
Baade -- Wesselink method to a higher precision must take into account
the structure and dynamics of Cepheid atmospheres, where the instantaneous
effective stellar radii corresponding to either the photometric or
interferometric version of the method do not precisely follow the
true motion of any particular layer of the atmosphere or velocities
found from Doppler distortions of any particular stellar line. Our
method of disentangling spectra provides a tool for observational
probing of the structure of the pulsating stellar atmospheres.

 Originally designed and well proved for disentangling of spectra
of multiple stars, our method can also be used for studies of Cepheids
in binaries, which can yield more information about the basic physical
parameters of the stars and thus also about the PL-relation in different
conditions. Further sophistications of the method are possible, e.g.
directly fitting the coefficients of the Fourier series given by
Eq.~(\ref{Fourser}) instead of independent velocities in individual
exposures to distinguish the orbital and pulsational changes of radial
velocities. These coefficients could be then used for the distance
determination in the way introduced by Krockenberger et al. (1997) or
for the classification according to Deb and Singh (2009).

 Another possible future improvement is to take into account for
the pulsational broadening functions the results of radiative transfer
and line formation in differentially pulsating atmospheres. This is
particularly challenging for non-radial pulsations, where the
approximation of an unperturbed moving atmosphere is even less physically
substantiated but even more commonly used to avoid the complexity
of the problem. Note that the output of residual spectra in rest
wavelengths of any component or the centre of mass of a multiple
system was implemented in the KOREL code by its author as a ``first
aid" for studying pulsational or other perturbations of spectra not
directly included in the model of the broadening functions $\Delta$.
However, such an approach should not be used as a black box without
understanding its underlying assumptions, namely that the free
parameters of the neglected effect (e.g. the pulsations) do not
correlate with the parameters taken into account (e.g. orbital
parameters). If we find a solution with a simplified model, it is
not proof that its assumptions are correct, because neglecting
a problem is not a true solution of the problem. The disentangling
of pulsations in spectra of single or multiple stars thus requires
one to include the pulsational variations in the model by
which we fit the observations. We have demonstrated such an approach here
in its simplest form. A future sophistication will be needed to account
for more accurate observations.


\begin{acknowledgements}
The authors thank the unknown referee and A. Han for useful comments.
This work has been done in the framework of the Center for Theoretical
Astrophysics (ref.~LC06014) with a support of grant GA\v{C}R 202/09/0772.
\end{acknowledgements}


\Online


\begin{thebibliography}{}


\bibitem[1926]{baa26} Baade, W. 1926, Astron. Nachr. 228, 359


\bibitem[2000]{ber00} Berdnikov, L. N. \& Ignatova, V. V. 2000, ASP Conf. Ser. 203, 244

\bibitem[1993]{bre93} Breitfellner, M. G. \& Gillet, D. 1993, \aap{} 277, 524

\bibitem[1993]{but93} Butler, R. P. 1993, \apj{} 415, 322

\bibitem[1996]{but96} Butler, R. P., Bell, R. A. \& Hindsley, R. B. 1996, \apj{} 461, 362

\bibitem[1997]{but97} Butler, R. P. \& Bell, R. A. 1997, \apj{} 480, 767

\bibitem[2009]{deb09} Deb, S. \& Singh, H. P. 2009, arXiv0903.3500

\bibitem[1991]{fok91} Fokin, A. B. 1991, \mnras{} 250, 258

\bibitem[2003]{fok03} Fokin, A. B. 2003, ASP Conf. Ser. 288, 491

\bibitem[1934]{get34} Getting, I. A. 1934, \mnras{} 95, 139


\bibitem[2005]{gra05} Gray, D. F. 2005, ``The observation and analysis of
 stellar photospheres" (Third edition), Cambridge University Press

\bibitem[2007]{gra07} Gray, D. F. \& Stevenson, K. B. 2007, \pasp{} 119, 398

\bibitem[1995]{had95} Hadrava, P. 1995, \aaps{} 114, 393

\bibitem[1997]{had97} Hadrava, P. 1997, \aaps{} 122, 581

\bibitem[1998]{had98} Hadrava, P. 1998, Proc. of the 29th conference on variable star research, HaP MK, Brno, p. 111

\bibitem[2003]{had03} Hadrava, P. \& Kub\'{a}t, J. 2003,
 ASP Conf. Ser. 288, 149 

\bibitem[2004]{had04a} Hadrava, P. 2004a,
 ASP Conf. Ser. 318, 86 

\bibitem[2004]{had04b} Hadrava, P. 2004b, Publ. Astron. Inst. ASCR 92, 15

\bibitem[2006]{had06a} Hadrava, P. 2006a, \aap{} 448, 1149

\bibitem[2006]{had06b} Hadrava, P. 2006b, \apss{} 304, 337

\bibitem[2009]{had09} Hadrava, P. 2009, \aap{} 494, 399



\bibitem[1993]{hil93} Hill, G. 1993, ASP Conf. Ser. 38, 127

\bibitem[2000]{kis00} Kiss, L. L. \& Vink\'{o}, J. 2000, \mnras{} 314, 420

\bibitem[2000]{kov00} Kovtyukh, V. V. \& Gorlova, N. I. 2000, \aap{} 358, 587

\bibitem[1997]{kro97} Krockenberger, M., Sasselov, D. D. \& Noyes, R. W. 1997, \apj{} 479, 875

\bibitem[1998]{kro98} Krockenberger, M., Sasselov, D. D. et al. 1998, ASP Conf. Ser. 154, 791

\bibitem[2002]{mar02} Marengo, M., Sasselov, D. D. et al. 2002, \apj{} 567, 1131

\bibitem[1985]{mof85} Moffett, T. J. \& Barnes, T. G. III 1985, \apjs{} 58, 843

\bibitem[2001]{mon01} Monta$\tilde{\rm n}$\'{e}s Rodriguez, P. \& Jeffery, C. S. 2001, \aap{} 375, 411

\bibitem[1966]{moo66}Moore, Ch. E., Minnaert, M. J. G. \&
 Houtgast, J. 1966, ``The solar spectrum 2935 {\AA} to 8770 {\AA}",
 National Bureau of Standards Monograph 61, Washington

\bibitem[2004]{nar04} Nardetto, N., Fokin, A. et al. 2004, \aap{} 428, 131

\bibitem[2006]{nar06} Nardetto, N., Mourard, D. et al. 2006, \aap{} 453, 309

\bibitem[2007]{nar07} Nardetto, N., Mourard, D. et al. 2007, \aap{} 471, 661

\bibitem[2008]{nar08} Nardetto, N., Kervella, P. et al. 2008, The Messenger 134, 20


\bibitem[2009]{nar09} Nardetto, N., Gieren, W. et al. 2009, \aap{} 502, 951 

\bibitem[2005]{pet05} Petterson, O. K. L., Cottrell, P. L. et al. 2005, \mnras{} 362, 1167

\bibitem[1971]{sch71} Schaltenbrand, R. \& Tammann, G. A. 1971, \aaps{} 4, 265


\bibitem[2002]{sko02}\v{S}koda, P. \& \v{S}lechta, M. 2002,
 Publ. Astron. Inst. ASCR 90, 22

\bibitem[1946]{wes46} Wesselink, A. J. 1946, \bain{} 10, 91

\bibitem[2008]{yan08} Yan, J., Liu, Q. \& Hadrava, P. 2008, \aj{} 136, 631

\bibitem[2006]{zuc06} Zucker, S. \& Mazeh, T. 2006, \mnras{} 371, 1513

\end{thebibliography}
\end{document}